# Study of genetic mutations and dynamic spread of SARS-CoV-2 pandemic and prediction of its evolution according to the SIR model

Hamieh Mohamad[1, a,b], Doumit Mary[1, a], Toufaily Joumana [2, a], Hamieh Tayssir[3, a]

[a] MCEMA and LEADDER Laboratories, Faculty of Sciences and EDST, Lebanese University, Hariri Campus, Hadath, Beirut, Lebanon.
[b] Engineering department, IUL University, Lebanon.

**Abstract**

**In this work, we aim to study that the dynamics behavior for cumulative number of SARS-CoV-2 pandemic can provide information on the overall behavior of the spread over daly time.The cumulative data can be synthesized in an empirical form obtained from a "Susceptible-Infected-Recovered" (SIR) model previously studied on a Euclidean network. From the study we carried out, we can conclude that the SIR model on the Euclidean network can reproduce data from several countries with a deviation of precision for given parameter values. This gives an insight into the different agents that influence the behavior of SARS-CoV-2 especially during the virus mutation period. We are thus trying to analyze the effect of genetic mutations in different countries, and how a specific mutation can make the virus more contagious.**

## Introduction

COVID-19 is the global crisis of our time and the biggest challenge we face these days. But this pandemic is much more than a health crisis, it is also an unprecedented socio-economic crisis. Putting pressure on each of the countries it affects, it has devastating social, economic and political impacts that will leave deep scars that will be slow to fade away [1, 2, 3].
The SARS-CoV-2 pandemic has spread at a phenomenal rate, this speed of propagation of the Sars-CoV-2 virus and its rapid evolution time leaves little time to react. By April 2020, the pandemic had already saturated the hospital capacities of many countries, causing more and more deaths around the world and seriously affecting the health of millions of people. The action time of the coronavirus is much faster than the different reaction times of human institutions, whether hospital, industrial or scientific research. Therefore, to understand the dynamics of the coronavirus pandemic, it is necessary to consider the timescale of virus action. The mathematical models that aim to predict this dynamic must take into consideration, the different transmission times of the virus and the clinical evolution of the disease that it causes. However, to cope with this emergency, scientists cannot only rely on complex computational models, whose implementation, responsiveness and data needs require weeks of collection and prohibitive computational costs. On the other hand, the importance of chronological aspects in understanding the pandemic potential of coronavirus could lead to less expensive mathematical models in terms of time calculation and data collection, especially when using real databases obtained on different tracks and in many countries. Thus, models that rely on systems of partial differential equations may not be adequate in the case of this pandemic. In addition, the age of the infected person does not appear to have a major role in the spread of the virus (at least according to current data). Only the evolution, severity and mortality rate of the disease strongly depend on it.

In consequency, the rapidity of propagation which is the pandemic strength of this virus could lead to simpler mathematical models, provided they take into consideration the chronological characteristics of the virus. Furthermore, one of the difficulties encountered in this pandemic study concerns the data collected, in particular those concerning the number of infected people. Many people contract the virus without reporting symptoms or severe forms of the disease, while they become contagious. In addition, the difficulty of carrying out virological tests in large numbers is pushing the refocusing of screening around cases that have developed severe symptoms. It is therefore not reasonable to base models only on the number of cases detected or confirmed. Even if, the proportionality of this indicator in relation to the total number of infected cases is not guaranteed, due to the evolution of the capacity to perform these tests, and the detection policy employed. On the other hand, one could easily admit that the comparison of the curve of evolution of transmission in different countries and populations, are markers which precisely reflect the evolution of the pandemic, and give remarkable clues on this evolution of the virus. Taking all these observations into account, we propose here a discrete deterministic evolution model over time and as a "one day" time interval. This evolution model is essentially based on the evolution of the cumulative number of patients. It is a decoupled model that makes it possible to simulate the evolution of various other indicators such as the propagation time, the change of rhythm of the curves over time and the different regimes observed by the cumulative growth curves. These various observations indicate predictions which are essentially based on the shape of the curves which makes the model insensitive to the onset of infection and then to the reaction rate of the populations confronted with this pandemic, in particular in the zigzag changes and the recurrence on the exponential law linked to the virus mutation. The model has a number of parameters, which are mainly the coefficients of susceptibility S, the new infected cases I, and the total cases over the number of populations in each country R. The time index gives an indication about the prediction of the continued virus propagation. The delays in time are deduced from the data obtained by the model used. As for the peak time obtained named "Tc", it is a dynamic marker that can be measured from the data and used to evaluate the impact of the measures applied. This marker has a fundamental role in predicting the pandemic pursue under different scenarios, and in the development of a technique in order to control it without maintaining permanent drastic measures [4, 5].

In this work, we propose a discrete model that allow us to follow day by day the evolution over time of the pandemic, taking into consideration the various medical parameters that characterize it; such as the virus incubation period, the outbreak and duration of contagiousness, the mean interval delay between the onset of symptoms, the social reaction and the virus mutation, etc. The model can be used to monitor the impact of restriction measures and the behavior of the population. It can also predict the short- and long-term dynamics of the pandemic, under several scenarios. In this case, the pandemic state will be particularly examined.

**Statistical study**

So far, statistics from reported cases show that over 80% of infected person had a mild case of the disease, while about 14% had severe symptoms, suffered from shortness of breath and pneumonia, and only about 5% are classified as patients with chronic disease, their symptoms include septic shock, respiratory failure, and multiple organ failure [6, 7, 8].

The original epicenter of the coronavirus was first identified in Wuhan, China, then the virus spread to other countries, most severely affecting Brazil, India, Russia and Europe. While the increase in the number of cases in the Western world is alarming, drastic precautionary measures have been taken in China. A clear picture of the spatiotemporal dependence of the spread of the virus can be best understood by analyzing the data from China, as the basic state for the virus rate is very broad. A considerable number of analyzes of the available data on the number of cases and deaths have been the subject of several recent studies, but few databases models have also been proposed [5, 9]. Notably, an exponential growth of time in early stage is noted in all the countries mainly affected; however, the number of deaths is considered to follow a law of force behavior [10]. This may be due to purely medical reasons; the possibility of survival also depends on the stage of detection and the treatment received.

We show in figure 1 the evolution of the virus over time for several countries such as France, Germany, Italy, Russia, USA, China, Lebanon, England and Iran. This evolution is taken in standardized mode in order to obtain a real comparison of the dynamic's propagation between these different countries. The exponential behavior is well obtained at the first stage of the pandemic, up to a period of approximately 70 days (similar for all the countries), this behavior remains applicable and allow to adjust the increasing curves. Beyond 70 days, a disturbance occurs on the exponential evolution which leads to an uncontrolled form. Some studies refer to this phenomenon for the containing of coronavirus cases [11]. We are therefore proposing a coherent strategy for controlling the pandemic, and getting out of total lockdown, without adopting a risky approach of the herd immunity type. This strategy, called the zigzag strategy, is based on the classification of the measures applied into four paths, distinguished by a marker called daily reproduction rate. The model and strategy in question are flexible and easily adaptable to new advances such as mass screenings or infection surveys. They can also be used at different geographic scales (local, regional or national).

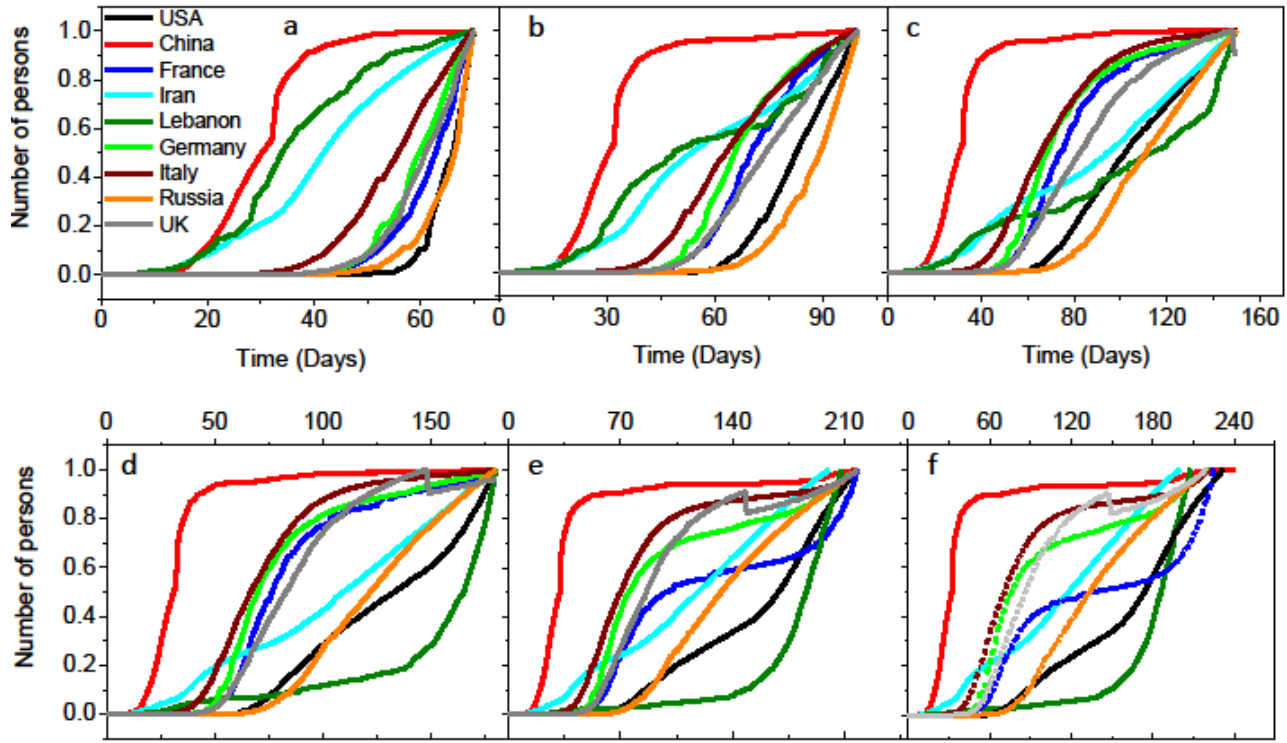

Figure 1: Standardization of COVID-19 outbreak for 9 countries (USA, China, France, Iran, Lebanon, Germany, Italy, Russia and UK). a) before the lockdown. b, c) after the lockdown. d) the transition from the first regime to the second regime. e, f) the emergence of the zigzag curve resulting from viral mutation.

At the same time, we notice that two regimes have evolved in the execution of this pandemic illustrated in figure (1.a), the first can be show by the dynamic evolution of the virus in China, Lebanon and Iran and so other countries. The curves follow an exponential law with a "Downward concavity" which increases up to 70 days, which can be explained by the accurate control of the zero point at the beginning of the contamination. Indeed, knowing the zero point and its origin allow us to control the evolution with a minimum processing time. This parameter remains an essential point for monitoring the evolution of the pandemic in the initial state. However, the second regime reported by France, Italy, Germany, Russia, USA and UK shows exponential growth with an increasing "Upward concavity" for the same 70-day period. This can be explained by the delay in reporting cases, as well as the zero point of infection, and consequently the cumulative numbers of patients remain almost discreet for a long time, which could influence the human reaction to fight against this pandemic, under this fold a strong growth is shown in the curves in a symmetric way at the first mode. These two regimes in the dynamics of evolution show a hysteresis behavior obtained by these two regimes.

Beyond 70 days (fig1-b), the behavior of the second regime begins to have a tendency of a positive "downward concavity", ie. In order, to switch to the first exponential regime it's necessary to wait up to 150 days (fig1-c). This is due to the natural behavior of the virus spreading to follow the same law obtained in China. However, an abnormality appears on the curves of Lebanon and Iran, as the passage occurs in the opposite direction, towards the second regime. This abnormality is explained by external parameters related to human social and humanitarian behavior as well as to the strategy followed to fight against the spread of Covid-19. Thus, some of these countries are going through a socio-economic bankruptcy which limits the reactions and the decisions for fighting against the virus propagation. Obviously, even if this deviance occurs, we subsequently notice that the behavior resumes to an exponential curve in order to continue its evolution, knowing that for the countries which have switched to the first regime, they have already gained time to stabilize the contamination for a certain period of time, compared to the other countries which must restart their transition again to the first regime such as Lebanon and Iran. Therefore, this indicates an extension of the time that will be added as an influence parameter for these countries indicated in figure (1-d). The data measurements beyond 150 days (fig1- e, f) show a zigzag behavior that we have discussed above. An explanation can also be linked to the zigzag shape, which could be caused by multiple mutations in the genetic code of the coronavirus, over time. This mutation is reproducible as shown by a proportional behavior of the curves between Italy, Germany, UK and France which has a dominant zigzag effect. On the other hand, it is clear that China, Italy, UK, Germany are currently in better shape in their evolution in the direction of slower growth, unlike USA, France, Russia, Iran and Lebanon which tend to evolve quickly.

**Adjustment of the experimental data model**
Since the virus can often be contracted once, therefore an affected person either dies or recovers. In this type of infectious diseases, the typical model that will be taken into consideration is: "Susceptible-Infected-Recovered" (SIR) type. In the SIR model, the population is divided into three categories: susceptible, infected and recovered; the total population (including deaths) is constant. As the disease spreads, susceptible people are likely to be become infected, then those infected they either die or recover (following this model, they are treated the same). The total infected population over time forms the deleted category. Usually the densities are considered and noted by S, I and R for the three categories and depend on the time t, with $I = dR/dt$ and $S + I + R = 1$.
In such pandemics: "I", corresponding to the newly infected density in a real situation, which initially shows a slow increase over time and which turns into a sharp increase before reaching a maximum value. Typically, the increasing phase shows exponential behavior. Once stability is reached, then we can declare that the pandemic is over. China is already in this phase of stability while most of the other countries have not yet reached the maximum value.
However, the cumulative data, "R", shows saturation once the peak value has been crossed. According to figure 2, we plot the data for the total fraction of R cases as a function of time for Italy, Iran and Russia. Thus, the insert shows the adjustment for China and USA where the total number of cases is divided by the total population N of each country compared to data obtained from several articles [12,13] where daily reports are available from January 21, 2020 (which means: day zero).

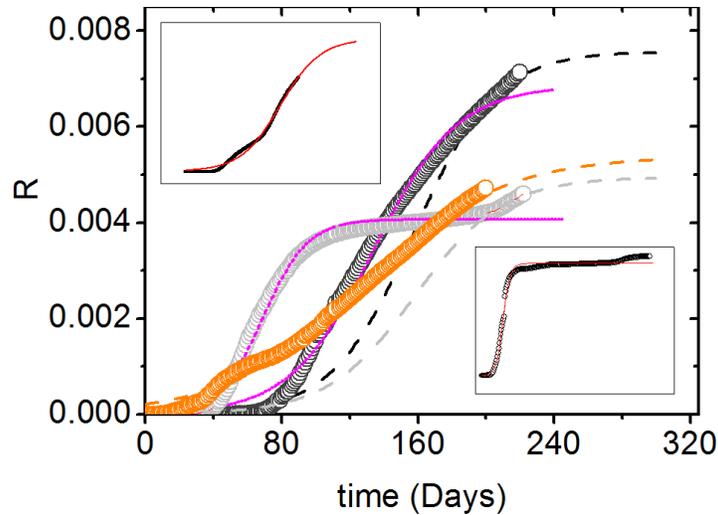

Figure 2: Adjustment of standardized data for the COVID-19 pandemic by SIR model. The black circles represent the curve of Russia. The gray circles represent the curve for Italy (the curve is similar to France, Germany and UK curves). The orange circles represent the Iranian curve.
In the upward insert, we have the adaptation of the US curve, while the downward insert shows the adaptation of the china curve.

We can empirically adjust the data to $R(t) = a \exp(t/T) / [1 + c \exp(t/T)]$ (1) with $a$, $c$ and $T$, all the parameters are shown in the table below (Table 1). The error involved for a and c is about 16% while for "T" it is 4.3%. This had already been noted up to day 40, in an earlier data report [14,15]. The data for the number of newly infected people are more dispersed and instead of a direct adjustment, they can be calculated by differentiating "R" from "t". The peak value of $I$ can be found by calculating the maximum of derivative $I = dR/dt$ and we can also locate the time $T_c$ where the peak occurred: $T_c = T \ln(1/c)$.

|  | a | c | T | $T_c$ |
|---|---|---|---|---|
| China | $1.2675 \times 10^{-7}$ | 0.00258 | 5.03 | 30 |
| Lebanon | $8.1708 \times 10^{-6}$ | 0.00296 | 37.71 | 219 |
| USA | $2.2 \times 10^{-4}$ | 0.00827 | 40.63 | 194 |
| France | $3 \times 10^{-5}$ | 0.00394 | 38.37 | 212 |
| Iran | $2.2 \times 10^{-4}$ | 0.04081 | 40.21 | 128 |
| Italy | $1.002 \times 10^{-5}$ | 0.00214 | 25.72 | 158 |
| Russia | $1.01 \times 10^{-5}$ | 0.00139 | 23.94 | 157 |
| Germany | $6.359 \times 10^{-6}$ | 0.002 | 24.88 | 154 |
| UK | $2.8855 \times 10^{-6}$ | 0.00053 | 21.07 | 158 |

Table 1: Represent the fitting parameters deduced from SIR model.

The adjustment curves represented in pink (fig. 2) show the real data according to the exponential law, and which is reproducible from one country to another. However, the offset obtained by the black dotted curve represents an adjustment according to the same law but after having changing the curve behavior of cumulative number over time. This change may be due to the mutations in the virus during the pandemic. The SARS-CoV-2 Spike protein is shown to be very susceptible to mutations, in which the number of mutations changes over time and differs between infected countries. Indeed, the explanation for this mismatch between the two adjustments may be related to these mutations' effects.

**Genetic analysis of SARS-CoV-2 and ACE2**

In fact, to understand the effect of the coronavirus mutations in the human body, we should firstly understand the encoded Angiotensin-converting enzyme 2 (ACE2) protein. This ACE2 is a functional receptor for the spike glycoprotein of the human severe acute respiratory syndrome coronaviruses, SARS-CoV and SARS-CoV-2 [16, 17]. The complete cDNA for human ACE2 encodes a protein of 805 amino acids. The human ACE2 is a type I integral membrane protein which contains a 17 amino acid N-terminal signal sequence and a 21 amino acid hydrophobic transmembrane sequence near the C-terminus followed by a 44 amino acid cytoplasmic domain, which contains potential phosphorylation sites and the predicted molecular weight of this protein is 85 kDa including tags [18]. Besides biochemical characterization, many 3D structures have been solved, for ACE2 protein, in total we have 21 ACE2 structures available in the Protein Data Bank (PDB) [19]. Based on the structures present in the PDB, we have several crystal structures, for example: (PDB code: 3scl) "Crystal structure of spike protein receptor-binding domain from SARS coronavirus epidemic strain complexed with human-civet chimeric receptor ACE2" may help to better see the interaction between ACE2 and SARS-CoV [20]. The contact between ACE2 and SARS-CoV-2 RBD (Receptor Binding Domain) can be divided into three regions. Previous study showed that the residues near lysine 31, and tyrosine 41, 82 - 84, and 353 - 357 in human ACE2 were important for the binding of S-protein in coronavirus [18, 21, 22].

A further studies and structure show how the receptor binding domain of SARS-CoV2 interacts with ACE2 and suggests that it is possible that two trimeric spike proteins bind to an ACE2 dimer [23]. There are many similarities of SARS-CoV-2 with the original SARS-CoV. Using computer modeling Xu et al [24] found that the spike proteins of SARS-CoV-2 and SARS-CoV have almost identical 3-D structures in the receptor-binding domain that maintains Van Der Waals forces. SARS-CoV spike protein has a strong binding affinity to human ACE2, based on biochemical interaction studies and crystal structure analysis [25]. SARS-CoV-2 and SARS-CoV spike proteins share 76.5% identity in amino acid sequences and, importantly, the SARS-CoV-2 and SARS-CoV spike proteins have a high degree of homology [24, 25].

Further analysis suggested that SARS-CoV-2 recognizes human ACE2 more efficiently than SARS-CoV, increasing the ability of SARS-CoV-2 to transmit from person to person [25]. Thus, the SARS-CoV-2 spike protein was predicted to also have a strong binding affinity to human ACE2 (table 2).

| Interaction between the RBD of SARS-CoV-2 & ACE2 | | |
|---|---|---|
| SARS-CoV | SARS-CoV2 | ACE2 |
| Arg$^{426}$ | Asn$^{439}$ | Tyr$^{41}$ Neutral, Polar |
| Tyr[484] → | Gln[498] Neutral, Polar side chain | Gln[42] Neutral, Polar side chain |
| | Thr[500] Neutral, Polar side chain | Lys[353] Basic, (+) charge |
| Thr[487] → | Asn[501] Neutral, Polar side chain | Arg[357] Basic, (+) charge |
| Val[404] → | Lys[417] Basic, (+) charge | Asp[30] Acidic, (-) charge |
| | Tyr[453] Neutral, Polar | His[34] Basic, (+) charge |
| | Gln[474] Neutral, Polar side chain | Gln[24] Neutral, Polar side chain |

Table 2: Several residues in the SARS-CoV-2 RBD, correspond to residue in SARS-CoV (some has been mutated, colored in red). The residues in SARS-CoV-2 can be recognized on the human ACE2 receptor and interact with them in several ways: hydrogen bonding, hydrophobic interactions [26].

In addition, from SARS-CoV RBD to SARS-CoV-2 RBD, the substitution of interface residues Tyr[442]→Leu[455], Leu[443]→Phe[456], Phe[460]→Tyr[473], Leu[472]→Phe[486] and Asn[479]→Gln[493] may also change the affinity for ACE2. SARS-CoV-2 RBD exhibited significantly higher binding affinity to ACE2 receptor than SARS-CoV RBD [26, 27, 28].

**Mapping the spread and the effects of SARS-CoV-2 Spike mutation D614G on transmissibility and pathogenicity**

Identifying the mutations in SARS-CoV-2 allows us to understand how the virus has evolved as it has traveled from population to population, for example: the types of SARS-CoV-2 viruses that we observe in tests from Europe and Asia, are different from the types we have seen in china and both are different from the types we're seeing in USA.
Mutations happen randomly and are part of the lifecycle, some mutations will break down the virus, Other mutations can benefit it, but so far, it's hard for a single letter mutation to lead to a different behavior of the virus. However, with the advancement of technology, it has become readily feasible to sequence the genome of the novel coronavirus and each person's coronavirus infection will yield a sequence of around 30,000 letters, which we can use these sequences to reconstruct which infection is connected to which mutation [29, 30].

According to some studies dated from June and July 2020, the Corona virus has been exposed to a genetic mutation that made it more contagious, but at the same time it lost its lethal effects. This study is based on a mutation referred to by the code "D614G", which has abounded in recent months and has become the dominant feature throughout the world. It is noted that before the date of last March 1, this mutation called "D614G" was present in only 10% of the genetic mutations undergone by the virus during its monitoring by researchers. It was named "D614G" because this mutation changes the amino acid at position 614, from -D- aspartic acid, to -G- glycine. As, we already know that the coronavirus is made of spike proteins, and, according to the researchers, "D614G" is also present in those same spikes. This mutation gives the virus more of these protrusions, and makes them more stable, which in turn facilitates their attachment to cells and their penetration [31, 32, 33].

But starting that month, this mutation was present in over 78% of genetic mutations. This mutation is considered very important because it affected the gene called Spike, and the protein of the same name, which is responsible for the concentration of the virus within the "ACE-2" receptor in human cells [34]. It's assumed, that these transformations allowed SARS-CoV-2 to spread densely and locate inside cells, which led to a clear increase in the pattern of infection, but at the same time it reduced its pathological symptoms. This means, in clearer terms, that the virus is spreading faster, but causing less damage. This mutation "D614G", and its strange spread among the viral samples that were examined, could explain the relatively strange numbers that were recorded in recent weeks in France, as the numbers of infected people witnessed a significant increase, and in contrast, the number of those who were hospitalized or were in the resuscitation department, there was no significant increase. Whereas, the number of French receiving intensive care on 19 August 2020 (after 200 days) reached nearly 374, compare to the 7,000 at the height of the pandemic. As we see in figure (3-b) which is also reproducible for Italy at the same point (200 days).

The high prevalence of the D614G mutation in Europe, North America and parts of Asia coincides with a decreasing in death rates from the coronavirus. Despite signs of diminishing severity of the virus, caution should continue to be exercised. As the explanation, which relies on the hypothesis that a genetic mutation has occurred, remains only a hypothesis, there is another explanation for this phenomenon is that the virus is now spreading among young age groups that have the ability to confront it. At the end, it should be noted that these mutations are unlikely to affect the effectiveness of future vaccines

**The rate evolution of the daily reproduction**

The daily reproduction rate have a key role in this model (explained below), especially in the reproduction rate of dynamic base. Keeping that in mind, the pandemic decreases until it is over. This rate is directly related to (1) the average number of contacts of infected persons on a given day, (2) the average probability of transmission and (3) the duration of contagiousness. The estimation from the pandemie data, allow to quantify the measures effectiveness taken and to predict realistic scenarios. On the curve we show the different dates of entry into total lockdown for the countries. A first observation could be made at this stage: the drastic measures of the complete lockdown lead to reproduction rates barely below of the threshold value 1. We note that, beside the drastic restrictions regime, the values are significantly greater than 1, even in periods with fairly large measurements. This observation suggests a difficulty in controlling the pandemic in the long term without the application of drastic measures. Thus, according to these estimation, even a moderate lockdown can only swing, the rate above 1.
In the case of SARS-CoV-2, we are facing a dilemma, (1) to maintain a total lockdown with numerous socio-economic damage or (2) to proceed with a lockdown that is not completely controlled with a risk of pandemic bouncing. We observe that in these cases the pandemic amplifies again and resumes its growth, with an average delay of 40 days, almost similar between the different countries (fig 3-a). This is well observed on the first derivative of the cumulative number, as a function of time for the standardized value of "I". Therefore, if the pandemic remains in symmetrical reproduction, i.e. up to 150 days reproduces up to 300 days, we could conclude that a third growth of the pandemic could be amplified and occurs after 300 days towards the end of November 2020 and the peak of the pandemic will reproduce again at the beginning of the year 2021.

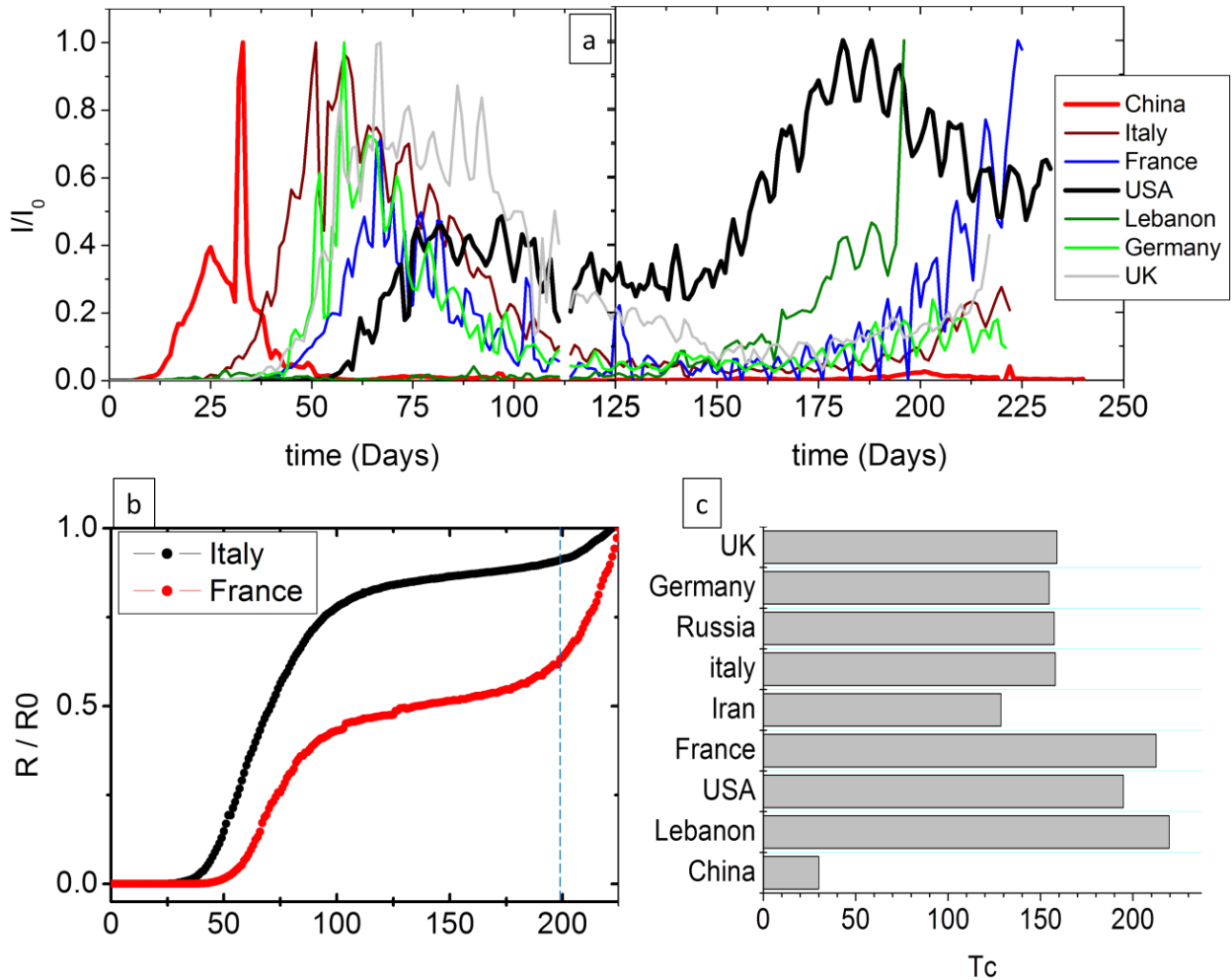

Figure 3: a-) The first derivative of the cumulative number as a function of time for the standardized value of "**I**". b-) Normalized **R** (dashed blue line represent 200 days). c-) A diagram describing the evolution of the pandemic over time in each country.

Another important element that results from this study is that the drastic measures are barely sufficient to explain the evolution of the pandemic in the countries studied. First, we note that we have a lot of flexibility to reduce these measures while controlling the pandemic. In other words, any exit strategy for lockdown, is a zigzag strategy, and remains a reproducible factor over time. However, the percentage of the population in each country and the periods of confinement, allow us to obtain a diagram describing the evolution of the pandemic over time. Figure (3-c) shows that Germany, Russia, Italy, UK, Iran remain less worrying than France, Lebanon and USA.

Finally, the model performed in this work can efficiently classify the measurement regimes applied in different countries. It results from all these elements of the evolution of the dynamic propagation of SARS-CoV-2 which passes through three stages: (1) The growth according to an exponential law with an inflection point which represents a natural behavior for this pandemic. (2) The unnatural behavior due to the social confinement imposed by human reaction in each society, this leads to a zigzag evolution not directly related to the lockdown system and which has the consequence of prolonging the infection time. (3) To confirm it according to the same exponential law to respect again the reproduction of the daily rate of infection. This may be linked in this study to the mutation of coronavirus in its global spread, which causes the pandemic to restart as a wave of propagation over time.